\documentclass[preprint,aps]{revtex4}
\usepackage{epsfig}
\usepackage{float}
\RequirePackage{amssymb,amsmath}
\makeatletter
\newcommand{\thickhline}{%
    \noalign {\ifnum 0=`}\fi \hrule height 1pt
    \futurelet \reserved@a \@xhline}
\usepackage[latin1]{inputenc}
\usepackage{graphicx}
\usepackage{indentfirst}
\usepackage{color}
\begin{document}

\title{Complete and incomplete fusion competition in
$^{11}$B-induced fission reaction on medium mass targets
at intermediate energies}

\author{N. A. Demekhina}

\affiliation{Joint Institute for Nuclear Research (JINR), Flerov Laboratory of Nuclear Reactions (LNR), Joliot-Curie 6, Dubna 141980, Moscow region Russia\\Yerevan Physics Institute, Alikhanyan Brothers 2, Yerevan 0036, Armenia\\E-mail: demekhina@nrmail.jinr.ru}

\author{G. S. Karapetyan}

\affiliation{Instituto de Fisica, Universidade de S\~ao Paulo, P. O. Box 66318, 05389-970 S\~ao Paulo, SP, Brazil\\E-mail: gayane@if.usp.br}

\author{A. R. Balabekyan}

\affiliation{Yerevan State University, Alex Manoogian St.1, 0025\\E-mail: balabekyan@ysu.am}

\begin{abstract}
The cross sections for the binary fission of $^{197}$Au, $^{181}$Ta and $^{209}$Bi targets induced by $^{11}$B ions were measured at intermediate energies. The fission products cross sections were studied by means of activation analysis in off-line regime observed $\gamma$-ray spectra. The fission cross section is reconstructed on the basis of charge and mass distribution of the fission products.

\end{abstract}

\maketitle

\section*{1 Introduction}
The heavy-ion-induced interaction with nuclei at energies near Coulomb barrier was intensively investigated the last decades. Particularly, the study of reaction mechanism concerning the transition between complete (CF) and incomplete fusion (ICF) processes is important in the view of understanding the interplay between such two dominant modes of the nuclear interaction. The complete fusion is experimentally measured value defined as a production of the nucleus with charge and mass equal to the sum of the masses and charges of a target and a projectile. When the projectile have been breakup near the target, several reaction mechanisms can be considered such as incomplete fusion, when one of the breakup fragments is captured by the target; and sequential complete or incomplete fusion, when breakup occurs followed by the successive capture of several or all the fragments by the target. The total fusion is defined usually as a sum of all processes resulting to the different interaction channels.   

The incomplete fusion is in generally theoretical value, based on the calculation of the different model approaches. The probability of the production of the particular reaction product can give an assumption about the contribution of the different reaction channels, which is frequently model depended. The investigation of the decay of weakly bound projectiles on the different components is very useful because allows to study of the breakup effect on the different reaction channels. The systematic experimental data in this range give the possibility to obtain the dependence on the charge and mass of both projectile and target for the nucleus-nucleus interaction in different energy range. The induced-activity method, used in present work, gives the information about distinct reaction channel. On the other site the off-line registration of the residuals yields cannot give the clear answer on the mechanism of its production. The reason is the following: one can measure the final reaction product usually after sequence of $\beta$-decays when the initial way of the production of giving nucleus is practically lost. Another reason is that by using the induced-activity method it is impossible to register stable or very short-lived isotopes. 
In present work we chose the fusion-fission channel of the interaction $^{11}$B-beam with $^{197}$Au, $^{181}$Ta and $^{209}$Bi at the incident energy of 24 AMeV as an indication of the fusion process resulting in target fission. Hereby, we tried to receive an information about fusion suppression process at energies above Coulomb barrier for the heavy elements, where fission may be comprised a noticeable part of the total reaction cross section. $^{11}$B represents a weakly bound system and can break up according to the following schemes: 

  $^{11}$B$\longrightarrow$$^{7}$Li + $^{4}$He    with  Q = -8.665 MeV;   
	
  $^{11}$B$\longrightarrow$$^{8}$Be + $^{3}$H$\longrightarrow$     with Q = -11.224 MeV
	
                  $\longrightarrow$$^{4}$He + $^{4}$He + $^{3}$H      with Q = -11.132 MeV.
									
   A capture of one of above mentioned contaminants of $^{11}$B would form a compound nuclei during incomplete fusion. The major competitive decay modes of this kind composed nuclei can be considered as:  $\sigma_{fis}$+$\sigma_{\alpha}$+$\sigma_{z=1}$+$\sigma_{xn}$ (fission, $\alpha$-particle, proton and x-neutrons emission). The measurements of the residual nuclei yields in the vicinity of target in these cases (xn and p+xn) cannot give the exactly determination of its production mechanism. For example, the production of the Tl isotopes in $^{197}$Au target cannot be separated from those produced by $\alpha$-particle absorption process after breakup of $^{11}$B, or from complete fusion of $^{11}$B-projectile and followed sequence $\alpha$-emission. Following $\beta$-decays can bring to the Hg isotopes which can be formed during other transformation also. On the other hand, the measurements of characteristic spectral $\gamma$-lines of different radioactive heavy elements such as Po, Bi, Pb, testify to fusion process or inelastic scattering resulting of transferred excitation energy and subsequent evaporation. 
Unfortunately the measurements of the induced activity in off-line regime do not allow separate the nuclei which concern to the same isobaric chain via $\beta$-decays. The excitation function of the different radioactive residues can give an information about its production mechanism also. The cross sections of the Tl-isotopes ($^{196-200}$Tl) increase with energy decrease, pointing out on the possibility of its production in different fusion processes, which is more probable at low energies. Neutron transfer reaction from the target or from the projectile with following production $^{196}$Au or $^{198}$Au isotopes are increased at lower energies, confirming the low energy character of these processes. Other Au isotopes ($^{195-191}$Au) do not reveal this behavior because of the cumulatively nature and contribution of the different channels in its formation. 
Taking into account all above mentioned aspects, in this paper we analyze the fission mode decay via production the fission fragments in mass range  $\sim$ 60-140 u from reactions $^{209}$Bi, $^{197}$Au and $^{181}$Ta with $^{11}$B-ions at the energies 255.5-137.5 MeV. The study of the fission cross section in the comparison with calculated data in frames of PACE-4 model can gives us hints about the probability of the compound nucleus production and its suppression at the intermediate energy range for the reason of the break up and incomplete fusion processes.

\section*{2 Experimental Procedure}

The irradiation  have been made on  U-400M accelerator on ACCULINA set up (JINR, Dubna Russia).The irradiation was made on the $^{11}$B-beam of energy 24 MeV/nucleon and 17 nA intensity during 12 hours. The targets along with Al-foils in a form of stack were placed normal to beam direction. The activities accumulated in targets were recorded using HPGe detectors. The residual radioactive nuclei were identified by the energy and intensity of characteristic $\gamma$-lines and by the respective half-lives of nucleus. Nuclear properties, used for identification of observed isotopes, were taken from literature \cite{Firestone}. The half-lives of identified isotopes were within the range of 1 h and 1 yr. The error in determining cross sections depended on the following factors: the statistical significance of experimental results ($\sim$ 2-3\%), the accuracy in measuring the target thickness and the accuracy of tabular data on nuclear constants ($\sim$ 3\%), and the errors in determining the detector efficiency with allowance for the accuracy in calculating its energy dependence ($\sim$ 10\%). The precision of the energy distribution depend on the energy scattering induced by the propagation throw step of the targets and Al absorbers.

In order to obtain a complete picture of the charge and mass distributions of fission products is necessary to estimate the cross sections of isotopes immeasurable by the induced-activity method.  It is necessary, therefore, to estimate the charge distribution curve (i. e., the variation of cross section with $Z$ at constant $A$) using independent cross section of the reaction products. Such variation can usually be expressed as a Gaussian distribution function \cite{Kudo}:

\begin{eqnarray}
\sigma_{A,Z}=\frac{\sigma_{A}}{(C\pi)^{1/2}}exp({-\frac{(Z-Z_{p})^{2}}{C}}),
\end{eqnarray}

\noindent where $\sigma_{A, Z}$  is the independent cross section for a given nuclide with an atomic charge $Z$ and a mass number $A$, $\sigma_{A}$ is the total isobaric cross section of the mass chain $A$, $Z_{p}$ is the most probable charge for that isobar, and $C$ is the width parameter of the distribution for the mass number $A$. Parameters of charge distribution determine the position of residue nucleus concerning stable isotopes with maximum yield in isobaric chain.

In the assumption of the constant width parameter of charge distributions ($C$) for different mass chains \cite{Kudo, Branquihno}, least-squares method was applied in order to get fitting parameters $Z_{p}$ and $\sigma_{A}$.

The mass distributions of the fission fragments have been analyzed in the frame of a two modes hypothesis, proposed by Brosa \cite{Brosa}, which consider existence of independent fission modes: symmetric ($S$) and asymmetric ($AS$). Mode S is mainly conditioned by the liquid drop properties of nuclear matter, and therefore the most probable values of fragment masses $A$ are close to $A_{f/2}$, where $A_{f}$ is the mass of the fissioning nucleus. The asymmetric mode AS with average masses of the heavy and light fragments $A_{H}$, $A_{L}$ with $Z_{H}$, $N_{H}$ and $Z_{L}$, $N_{L}$ (proton and neutron numbers of a heavy and light fragments) close to the one of the shell numbers.
In the multimodal model, the fission cross section as a function of mass number is obtained by the sum of three Gaussian functions, corresponding to symmetric and asymmetric fission modes \cite{Younes}:

\begin{align}
 \begin{split}
  Y_A=&
\frac{1}{\sqrt{2\pi}}\bigg[\frac{K_{AS}}{\sigma_{AS}}
\exp\left(-\frac{(A-A_S-D_{AS})^2}{2\sigma^2_{AS}}\right)+
\frac{K'_{AS}}{\sigma'_{AS}}\exp\left(-\frac{(A-A_S+D_{AS})^{2}}
{2\sigma'^2_{AS}}\right)+\\
&\frac{K_S}{\sigma_S}\exp\left({-\frac{(A-A_S)^2}
{2\sigma^2_S}}\right)
\bigg],
 \end{split}
\end{align}
where $A$ is the fragment mass number; $A_S$ is the mean mass number 
which determines the center of the Gaussian functions; and $K_i$, $\sigma_i$, 
and D$_i$ are the contribution, dispersion and position parameters of 
the $i^{th}$ Gaussian functions. The indexes $AS$ and $S$ designate the 
asymmetric and symmetric components.

\section*{3 Results and Discussion}

The experimental cross sections of fission fragment production in the mass range of 59 $\leq A \leq$ 136 u formed by the interaction of $^{11}$B-ions with Au,  Bi and Ta targets in the energy range 255.5 - 137.5 MeV are presented in Figs. 1-3, respectively. Integrating over the Gaussians and multiplying with a factor 0.5, because of the two fission fragments in each fission event, gives an estimate for the fission cross section. In Figs. 1-3 the mass-yield distributions, obtained by fitting procedure, are represented by the solid curves. From the obtained mean values of the fragment mass distributions it can be concluded that a mass of $\sim$ 194 u is expected for the fissioning nucleus in the case of gold target at the projectile energy 255.5 MeV, respectively. For the system $^{11}$B+$^{209}$Bi we expect $A_{f}$ $\sim$ 208 at the energy 146 MeV u and for Ta target at 245.4 MeV the average fissioning nucleus has been estimated as $\sim$ 177.4 u.

In analysis the fission process of $^{209}$Bi we were taking into account the contribution of symmetric and asymmetric parts in the distribution of the fission fragments. The values of the fission cross section studied in present experiment are contained in Table I. For comparison in the same place are included also the calculated fission cross section by fusion-evaporation code PACE-4 \cite{Tarasov} in assumption of complete fusion of $^{11}$B-ions. The calculated fission cross sections essentially exceed the experimental ones. From the comparison of experimental and calculated fission cross section, one can see that the suppression of the complete fusion process in the energy range under study varies from $\sim$ 25 \% to $\sim$ 30 \%. In the study of fusion suppression for $^{9}$Be + $^{144}$Sm system \cite{Gomes} the authors have proposed the dependence of the incomplete fusion on the target charge through the more probable breakup process with increasing the target charge.  It was expected also that at intermediate energy range the breakup probability is proportional to the gradient of the nuclear potential multiplied by an exponential factor depended of surface to surface separation. According to the dynamic model \cite{Wilczynski}, heavy-ion induced fusion-fission reactions at intermediate energy are characterized by the formation of a composite nucleus with high excitation energy and large angular momentum. Calculated fission-barrier heights as a function of angular momentum have shown a lowering with increasing of angular momentum and fission is expected to play a significant role. It can be suggested that the experimental fission cross sections decreasing mainly results of interaction via incomplete fusion process with different parts of the projectile (as $^{4}$He, $^{7}$Li, $^{8}$Be, $^{3}$H-ions) because the fission cross sections induced by these projectile fragments have essentially lower values. From our experiments these properties are revealed in connection with targets fissionability. For Au target the fissionability difference in the interaction with $^{11}$B and its different breakup components as $^{4}$He, $^{3}$H, $^{7}$Li nuclei is sufficiently. The fission cross sections for $^{197}$Au by $\alpha$-particles at energy 140 MeV \cite{Buttkewitz} is equal about 100 mb, for $^{181}$Ta this value at the same energy range is low also ($\sim$ 10 mb) \cite{Vigdor}. In reactions induced by $^{7}$Li ions the cross sections do not expected large based on the data from $^{197}$Au + $^{6}$Li reactions \cite{Vigdor}. The same results are for $^{181}$Ta fission using the same projectiles \cite{Vigdor}. By such a reason for above mentioned targets the picture is approximately the same including incomplete fusion suppression concerning of the fission process. On the other hand, the experimental data for $^{209}$Bi \cite{Dasgupta, Hassan} shows a sufficiently large fission cross section in reactions induced by $^{4}$He and $^{7}$Li ions in energy range studied in this experiment. Therefore for $^{209}$Bi target a lower effect of incomplete fusion process can be expected and accordingly the lower suppression of the complete fusion ($\sim$17 \%) was observed.

\section*{4 Conclusion}

The present data fill a gap of data concerning fission fission cross section for heavy-ion-induced fission at intermediate energies.
The fragment mass distribution in the case of gold target at intermediate energies shows Gaussian mass distributions.
The increase in the relative width of the mass distribution

\medbreak
\begin{center}
Table I. Experimental ($\sigma_{EXP}$) and calculated fusion-fission cross section ($\sigma_{CF}$) as well as suppression of the complete fusion cross section:  ($\sigma_{CF}$-$\sigma_{EXP}$)/$\sigma_{CF}$ (\%).
\end{center}
\begin{center}
\begin{tabular}{c c c c c}
\thickhline
Target& Energy, MeV& $\sigma_{EXP}$, mb& $\sigma_{CF}$, mb& $\sigma_{CF}$-$\sigma_{EXP}$)/$\sigma_{CF}$ (\%)\\ \hline
$^{209}$Bi& 146$\pm$15 MeV& 1263.6$\pm$130 MeV& 1520& 17$\pm$3\\ 
$^{197}$Au& 255.5$\pm$7 MeV& 785$\pm$80 MeV& 1210& 35$\pm$4\\ 
$^{181}$Ta& 245.4$\pm$9 MeV& 35.8$\pm$7 MeV& 52.7& 32$\pm$3.5\\ \hline
\end{tabular}
\end{center}
\vspace{2cm}

\newpage
\begin{figure*}[h!]
\includegraphics[width=16cm]{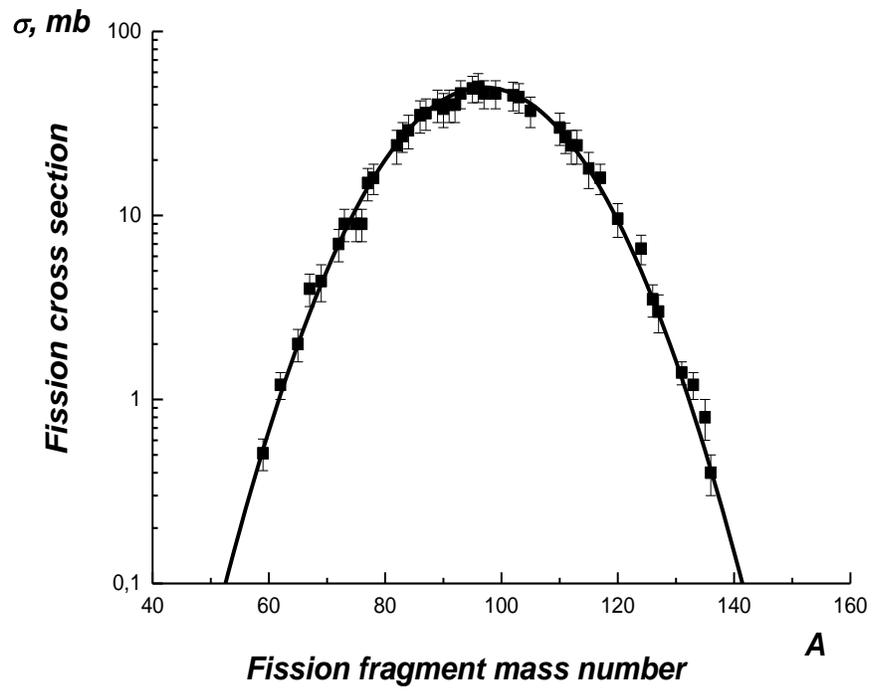}
\caption{\small Mass-yield distribution of fission fragments for $^{197}$Au target at 255.5 MeV.}
\end{figure*}

\newpage
\begin{figure*}[h!]
\includegraphics[width=16cm]{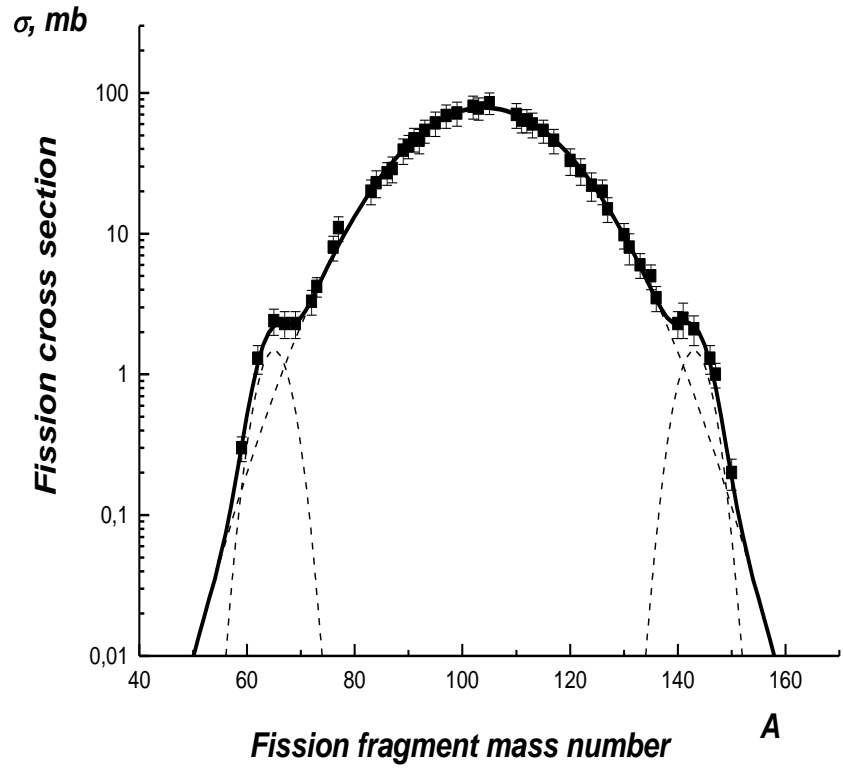}
\caption{\small Mass-yield distribution of fission fragments for $^{209}$Bi target at 146 MeV.}
\end{figure*}

\newpage
\begin{figure*}[h!]
\includegraphics[width=16cm]{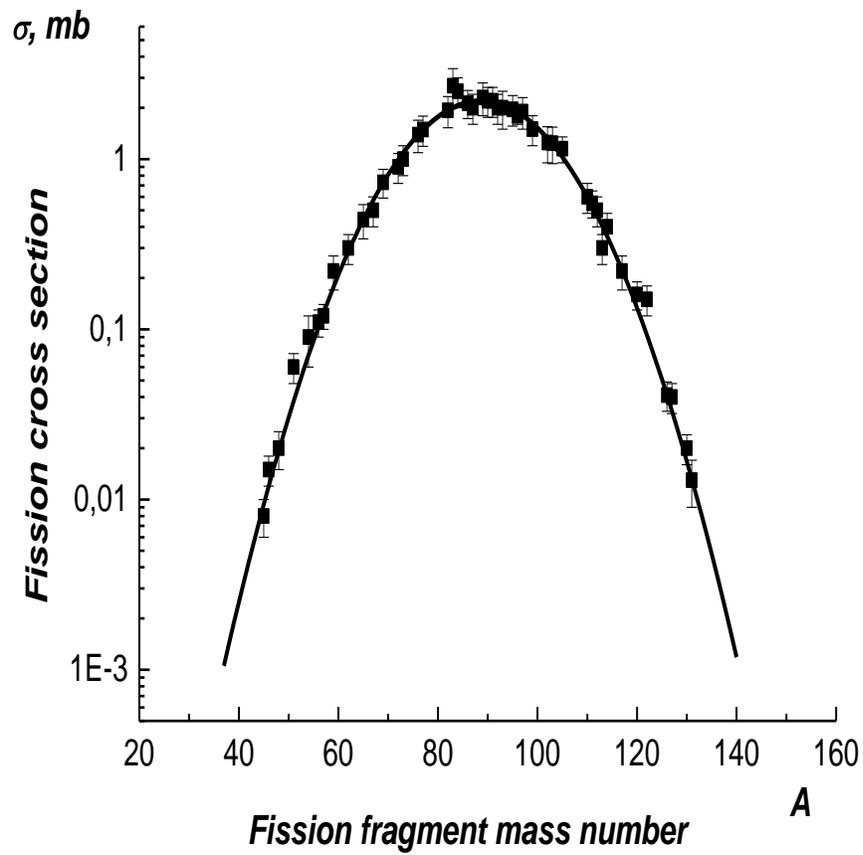}
\caption{\small Mass-yield distribution of fission fragments for $^{181}$Ta target at 245.4 MeV.}
\end{figure*}

\section*{Acknowledgment}
G. Karapetyan is grateful to Fundacao de Amparo a Pesquisa do Estado de Sao Paulo (FAPESP) 2011/00314-0 and 2014/00284-1, and also to International Centre for Theoretical Physics (ICTP) under the Associate Grant Scheme.

\medbreak
\end{document}